\documentclass[pra, twocolumn,showpacs,superscriptaddress,groupedaddress, noshowpacs]{revtex4}  % for review and submission
\usepackage{graphicx}  % needed for figures
\usepackage{dcolumn}   % needed for some tables
\usepackage{bm}        % for math
\usepackage{amssymb}   % for math
\usepackage{float}
\usepackage{mathtools}
\usepackage{mathrsfs}
\usepackage{amsmath}
\usepackage{bbm}
\usepackage{dsfont}

\newcommand{\ket}[1]{\ensuremath{|#1\rangle}}

\newcommand{\cPh}{\hat{\begin{cal}P\end{Huge}{cal}}}

\usepackage{ulem} %for \sout
\usepackage{color}

\begin{document}

\title{Transport of atoms across an optical lattice using an external harmonic potential}

\author{Tom Dowdall}
\address{Department of Physics, University College Cork, Cork, Ireland}

\author{Andreas Ruschhaupt}
\address{Department of Physics, University College Cork, Cork, Ireland}

\begin{abstract}

Precise control of quantum particles is required for many interesting or novel experiments. Here we consider the task of transporting an atom using an external harmonic potential from one well of an optical lattice to another without motional excitations. To achieve this we apply techniques from Shortcuts to Adiabaticity (STA) enabling fast and robust state manipulation. The process is split up into three independent building blocks; first the atom is loaded into an additional external harmonic trap; this trap is then transported from one lattice site to another and finally the atom is unloaded back onto the lattice by opening the external harmonic trap. We design protocols for each of these building blocks separately using invariant-based inverse engineering. Additionally we extend this method to the transport of a Bose-Einstein condensate described by the Gross-Pitaevskii equation.
\end{abstract}
\maketitle

\section{Introduction}
Robust high fidelity control of quantum systems is essential for all quantum technologies. Of particular interest is the movement of particles without motional excitations. Optical tweezers have become a common approach to enable precise control of single atom experiments and in recent years have been used to atom-by-atom assemble arrays in two and three dimensions \cite{barredo_1,barredo_2}. A major application of these optical tweezers has been as a means of transporting particles \cite{Beugnon_Two_dimensional_transport} and trying to achieve robust and lossless transport on shorter than adiabatic timescales \cite{Couvert_Optimal_Transport_OT}.

Other applications of optical tweezers are used to assemble defect-free one-dimensional arrays of cold neutral atoms \cite{manuel_edres}, motivated by a number of applications such as many qubit experiments or studying many-body physics in the Hubbard model, such as antiferromagnetic spin chains in an optical lattice \cite{Murmann_Anti_spin} or entangling neutral atoms using local spin exchange \cite{Kaufmann_entangling_local_exch}.

The ability to manipulate arrays of atoms on a lattice immediately has applications of realizing Maxwell's demon in a three-dimensional lattice \cite{Kumar_maxwell}. Here the sorting of a lattice, such that every site is filled, leads to a lower entropy state. This has potential as a first step towards neutral atom quantum computers. There are also applications, in the manipulation of Bose-Einstein condensates for mixing different species \cite{HCN_Condensate_mixing}  for experiments in many body quantum physics such as Bose polarons created through impurities in condensates \cite{Hu_Bose_polaron}. Both the transport and loading of atoms are important ingredients in all these experiments and applications.

To prepare and manipulate all these quantum systems, fast and robust protocols are required. A typical approach to manipulate these quantum systems is through the use of an adiabatic Hamiltonian; however this Hamiltonian must be varied sufficiently slowly to avoid excitations \cite{messiah_book}. Adiabatic processes have long process times and are vulnerable to decoherence; this makes them unsuitable for processes that need to be both fast and robust.

One set of techniques to achieve a more robust manipulation is Shortcuts to Adiabaticity \cite{sta_review_1,sta_review_new}. This collection of techniques allows for high fidelity preparation and manipulation of quantum systems on short time-scales. Previous works have demonstrated the effectiveness of Shortcuts to Adiabatacity for transport of particles \cite{Torrentegui_transport} along with fast trap variations \cite{chen_2010a} and have extended this treatment to Bose Einstein condensates \cite{bec_transport,bec_trap_var}.

In this paper we will develop schemes to transport atoms across an optical lattice using  techniques from Shortcuts to Adiabaticity. We examine a number of different strategies for achieving fast and robust transport of atoms or Bose-Einstein condensates over a lattice using invariant engineering.

In Sect. II we will lay out the model of the lattice and external potential system and will break down the transport process into three building blocks. In Sect. III we will then develop shortcut schemes for these different building blocks to enable fast and robust transport. 
\section{Model and Shortcuts}
\begin{figure}[t]
	\begin{center}
	(a)
		\includegraphics[width = 0.6 \linewidth]{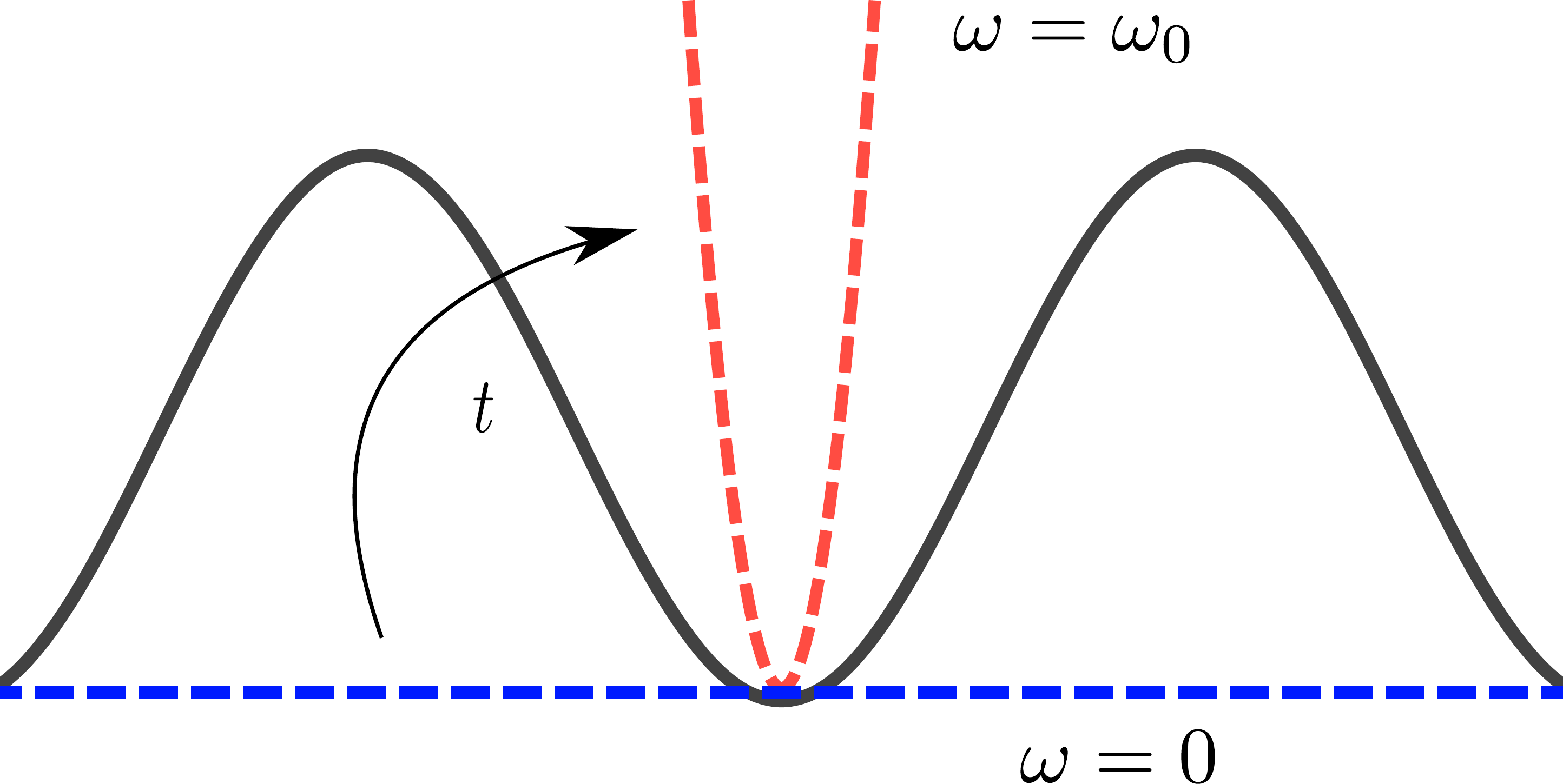}\\
	(b)
		\includegraphics[width = 0.6 \linewidth]{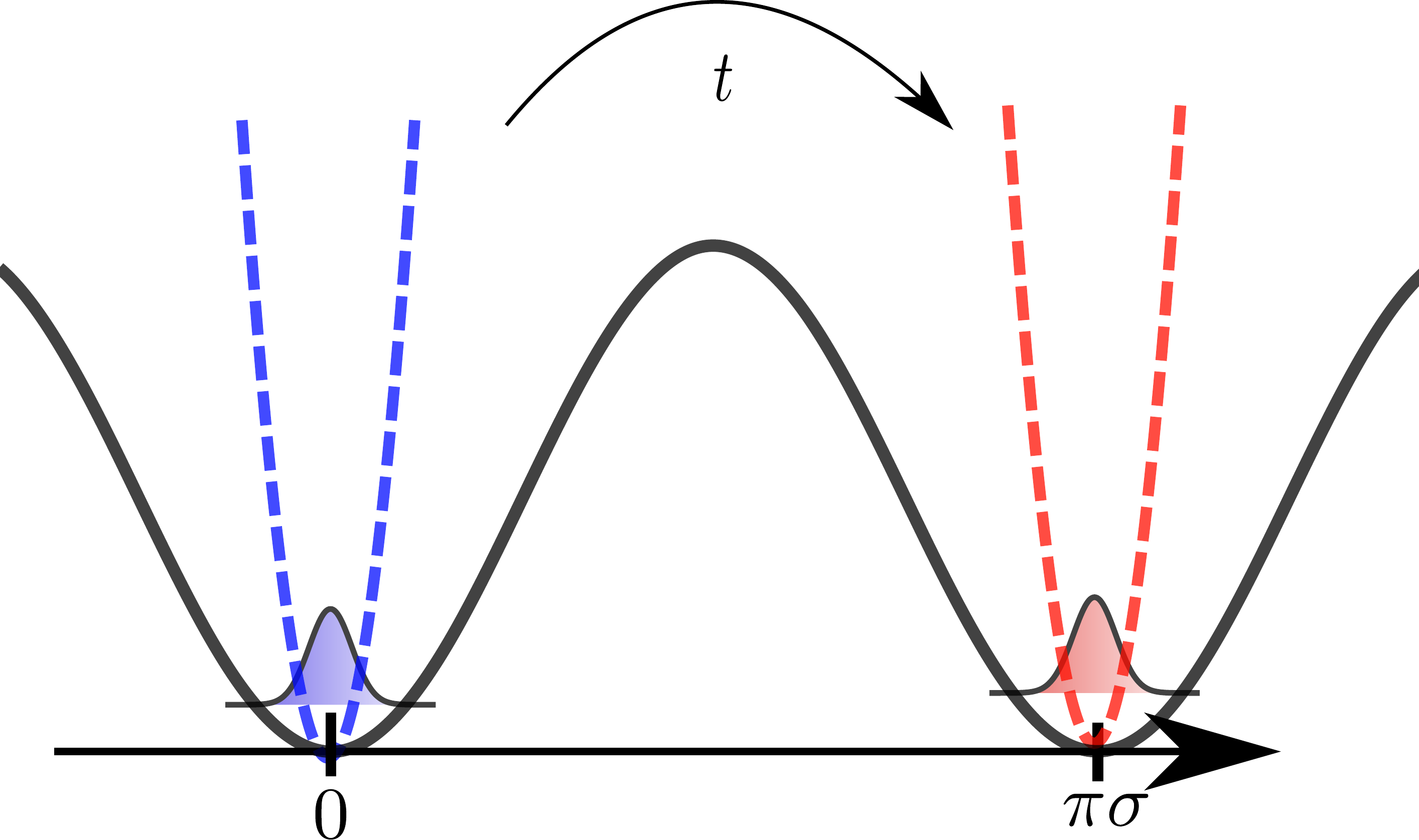}\\
	(c)		
		\includegraphics[width = 0.6 \linewidth]{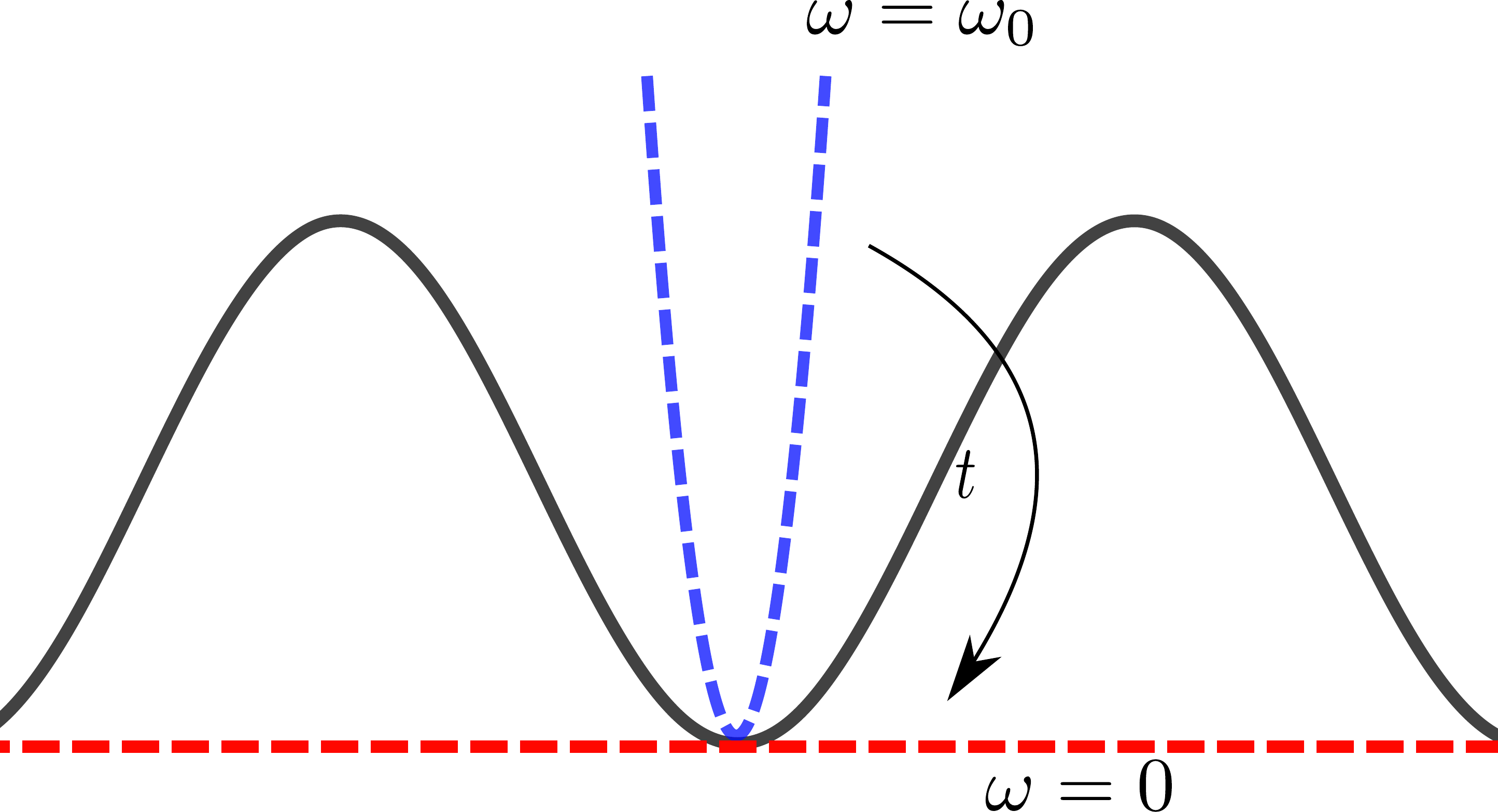}
	\end{center}
	\caption{Schematic of different building blocks: (a) loading an atom or condensate into an external harmonic trap, (b) shifting the trap across a lattice, (c) unloading the atoms or condensate back onto lattice.}
	\label{fig:sche_unloading}
\end{figure}

We consider a potential consisting of an external harmonic trap and an optical lattice in one dimension; the potential $V(x,t)$ of such a system is given by,
\begin{align}
V(x,t) = \dfrac{1}{2} m \omega(t)^{2} (\hat{x} - q_{0}(t))^{2} + U_{0} \sin^{2}\left(\dfrac{\hat{x}}{\sigma}\right)
\label{eq:fullpot}
\end{align}
where the trap frequency $\omega(t)$ and the trap centre position $q_{0}(t)$ are time dependent. Let us start by considering a single quantum particle governed by the one dimensional Schr\"odinger equation,
\begin{equation}
i \hbar \dfrac{\partial}{\partial t} \ket{\psi(t)} = \bigg[\dfrac{p^{2}}{2m} + V (x,t)\bigg] \ket{\psi(t)}.
\end{equation}
Further we also discuss a Bose-Einstein condensate governed by the Gross-Pitaevskii equation,
\begin{eqnarray}
i \hbar \dfrac{\partial}{\partial t} \ket{\psi(t)} = \bigg[ \dfrac{p^{2}}{2m} + V(x,t) + g(t) |\psi(x,t)|^{2} \bigg] \ket{\psi(t)}.
\end{eqnarray}
The $g(t)$ here models the atom-atom interaction in the condensate.\\
\\
Our goal is to transport a particle or Bose-Einstein condensate from one lattice site to another using the external trap. To achieve this we split the transport process into three building blocks as follows:
\begin{enumerate}
	\item Loading particles initially on a lattice into the external harmonic trap see Fig. \ref{fig:sche_unloading}(a)
	\item Shifting of particles confined in the external harmonic trap across an optical lattice see Fig. \ref{fig:sche_unloading}(b)
	\item Opening the harmonic trap and unloading the particles back into the lattice see Fig. \ref{fig:sche_unloading}(c).
\end{enumerate}
 Through the concatenation of these steps we can move particles across many different lattice sites. We will apply STA techniques to design each of these building blocks to achieve fast and robust movement across the lattice.
%
%
% ------------------------ Fig. 1 -------------------------------
% ---------------------------------------------------------------
%
%
\subsection{Invariant-based Inverse Engineering}
To derive a shortcut scheme for the single particle case we first make a harmonic approximation of the potential in Eq. (\ref{eq:fullpot}) obtaining
\begin{eqnarray}
V(x,t) = \dfrac{1}{2} m \tilde{\omega}(t)^{2} (\hat{x}-x_{min}(t))^{2} + V(x_{min}),
\label{eq:harm_approx}
\end{eqnarray}
where we have the frequency of the virtual harmonic trap $\tilde{\omega}(t)$ and trap centre position for the virtual trap $x_{\text{min}}(t)$ related to the real trap frequency $\omega(t)$ and real trap centre position $q_{0}(t)$ by
\begin{equation}
\omega(t)^{2} =  \tilde{\omega}(t)^{2} - \Omega^{2} \cos\left(2 x_{\min}(t)/\sigma\right),
\end{equation}
\begin{equation}
q_{0}(t) = x_{min}(t) + \dfrac{\Omega^{2}}{\omega^{2}} \sin\left(\dfrac{x_{\text{min}}(t)}{\sigma}\right)\cos\left(\dfrac{x_{\text{min}}(t)}{\sigma}\right).
\label{eq:q0(t)}
\end{equation}
We have defined the frequency $\Omega = \sqrt{2 \dfrac{U_{0}}{\sigma^{2} m}}$; this $\Omega$ corresponds to the frequency of the harmonic approximation of the well of the lattice. We will also use a time unit $T$, defined by $T = 1/\Omega$. \\
\\
Now that we have this system approximated by a harmonic trap we can apply the analysis developed in \cite{chen_2010a,Torrentegui_transport} to develop shortcut schemes for it. We want to start in an eigenstate of the Hamiltonian
\begin{equation}
H(t) = \dfrac{\hat{p}^{2}}{2m} + \dfrac{1}{2} m \tilde{\omega}(t)^{2} (\hat{x}-x_{min}(t))^{2}
\label{eq:hamil}
\end{equation}
at initial time $t=0$ and finish in an eigenstate of the the final Hamiltonian $t=t_{f}$, with the external harmonic trap shifted over a lattice site or with loading or unloading into the lattice. To do this we use the method of inverse engineering, using the Lewis-Reisenfeld invariant \cite{lr_invariant}. The Hamiltonian in Eq. (\ref{eq:hamil}) has a dynamical invariant
\begin{align*}
I(t) = \dfrac{1}{2m} \left[ \rho (\hat{p} - m q_{c}) - m \dot{\rho} (\hat{x} - q_{c})\right]^{2}\\
+ \dfrac{1}{2} m \tilde{\omega}(0)^{2} \left(\dfrac{\hat{x} - q_{c}}{\rho}\right)^{2}.
\label{eq:Invariant}
\end{align*}
The $\rho(t)$ and $q_{c}$ functions are auxiliary functions that have to obey auxiliary equations
\begin{equation}
\rho^{3}(\ddot{\rho} + \rho \tilde{\omega}^{2})  - \tilde{\omega}_{0}^{2} = 0,
\label{eq:Aux_eqns_1}
\end{equation}
\begin{equation}
\ddot{q}_{c} + \omega^{2}(q_{c}-x_{\text{min}}) = 0.
\label{eq:Aux_eqns_2}
\end{equation}
The Eqs. (\ref{eq:Aux_eqns_1}) and (\ref{eq:Aux_eqns_2}) relate the auxiliary functions $\rho(t)$ and $q_{c}$ to the virtual harmonic trap parameters $x_{\text{min}}(t)$ and $\tilde{\omega}(t)$.

To derive the appropriate boundary conditions on $\rho(t)$ and $q_{c}(t)$ for high fidelity state transition, we demand that the Hamiltonian $H(t)$ and the invariant $I(t)$ commute at initial and final times \textit{i.e.}  $[I(0),H(0)] = [I(t_{f}),H(t_{f})]=0$. From the resulting expressions, we obtain the boundary conditions on the auxiliary functions $\rho(t)$ and $q_{c}(t)$,
\begin{eqnarray*}
\rho(0) = 1  ;&q_{c}(0)= x_{min}(0);\\
\rho(t_{f}) = \sqrt{\dfrac{\tilde{\omega}(0)}{\tilde{\omega}(t_{f})}};& q_{c}(t_{f}) = x_{min}(t_{f}); \\
\dot{\rho}(0) = 0 ;&\dot{q}_{c}(0) = 0;\\
\dot{\rho}(t_{f}) = 0 ;&\dot{q}_{c}(t_{f})= 0.
\end{eqnarray*}
We further set boundary conditions on the second derivatives  $\ddot{\rho}(0) = \ddot{\rho}(t_{f}) = 0$ and $\ddot{q}_{c}(0) = \ddot{q}_{c}(t_{f}) = 0$ to ensure smoothness of the control functions $\omega(t)$ and $q_{0}(t)$. This approach is extended to the case of Bose-Einstein condensates in Appendix \ref{BEC_extension}. In the case of Bose-Einstein condensates, we obtain a further auxiliary equation 
\begin{align}
g(t) = \frac{g_{0}}{\rho(t)}.
\label{eq:Aux_eqns_3}
\end{align}
Feschbach resonance can be used to tune the atom atom interaction $g(t)$ according to Eq. (\ref{eq:Aux_eqns_3}). We can now fix the functions $\rho(t)$ and $q_{c}(t)$ according to the boundary conditions, then inverting the auxiliary equations Eq. (\ref{eq:Aux_eqns_1}) and Eq. (\ref{eq:Aux_eqns_2}) to obtain
\begin{align}
\tilde{\omega}(t)^{2} = - \dfrac{\ddot{\rho}(t)}{\rho(t)} + \dfrac{\tilde{\omega}(0)^{2}}{\rho(t)^{3}}
\label{eq:w_control}
\end{align}
\begin{align}
x_{min}(t) = q_{c}(t) + \dfrac{\ddot{q}_{c}(t)}{\tilde{\omega}(t)^{2}}.
\label{eq:x_control}
\end{align}
Now we proceed to develop schemes for the different building blocks using shortcuts framework.
\section{Building Blocks of Transport}
\subsection{Loading particles into a harmonic trap}
\label{subsec:loading}
\begin{figure}
	\begin{center}
	(a)\\
		\includegraphics[width = 0.8 \linewidth]{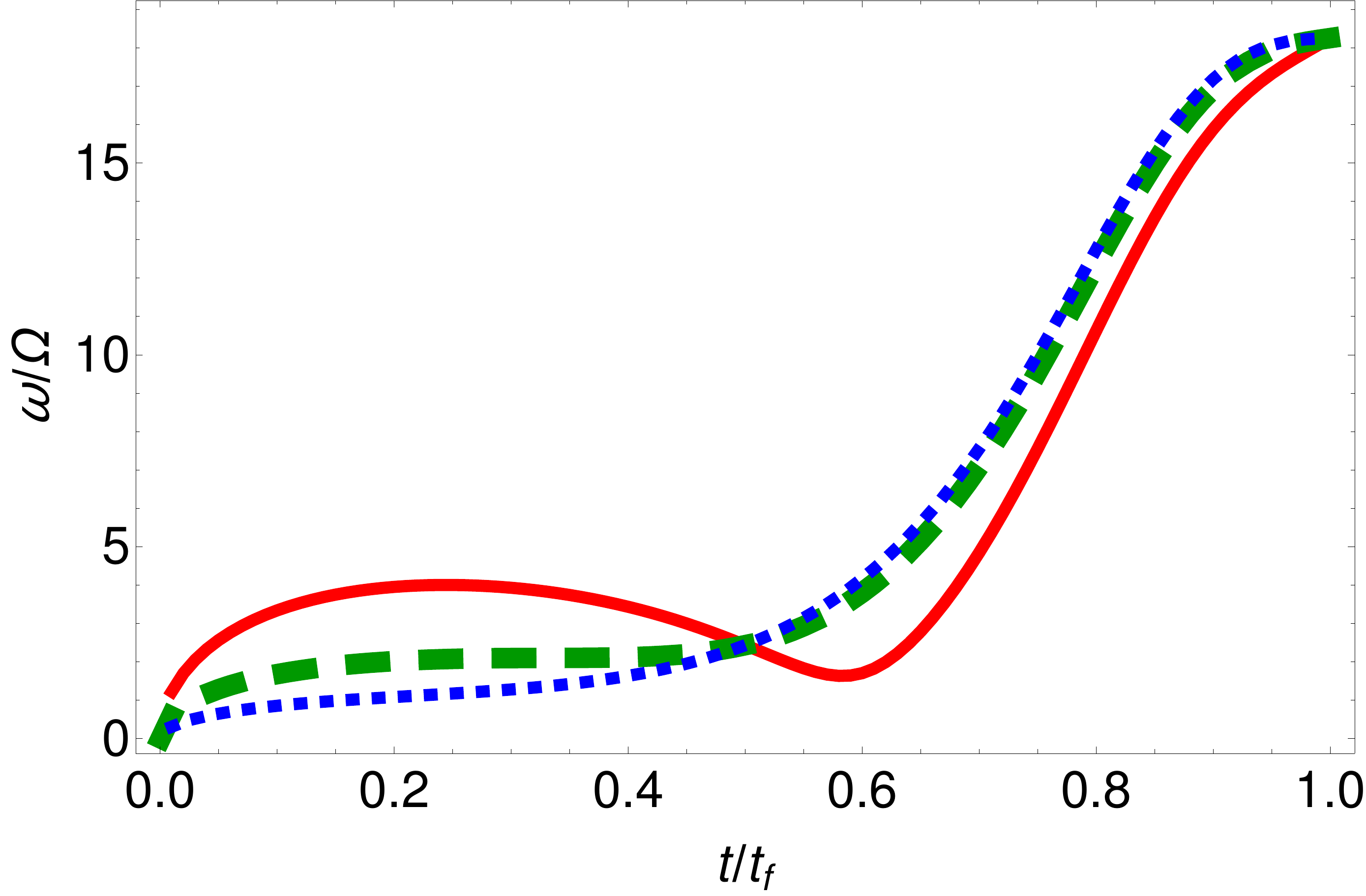}\\
		(b)\\		
		\includegraphics[width = 0.8 \linewidth]{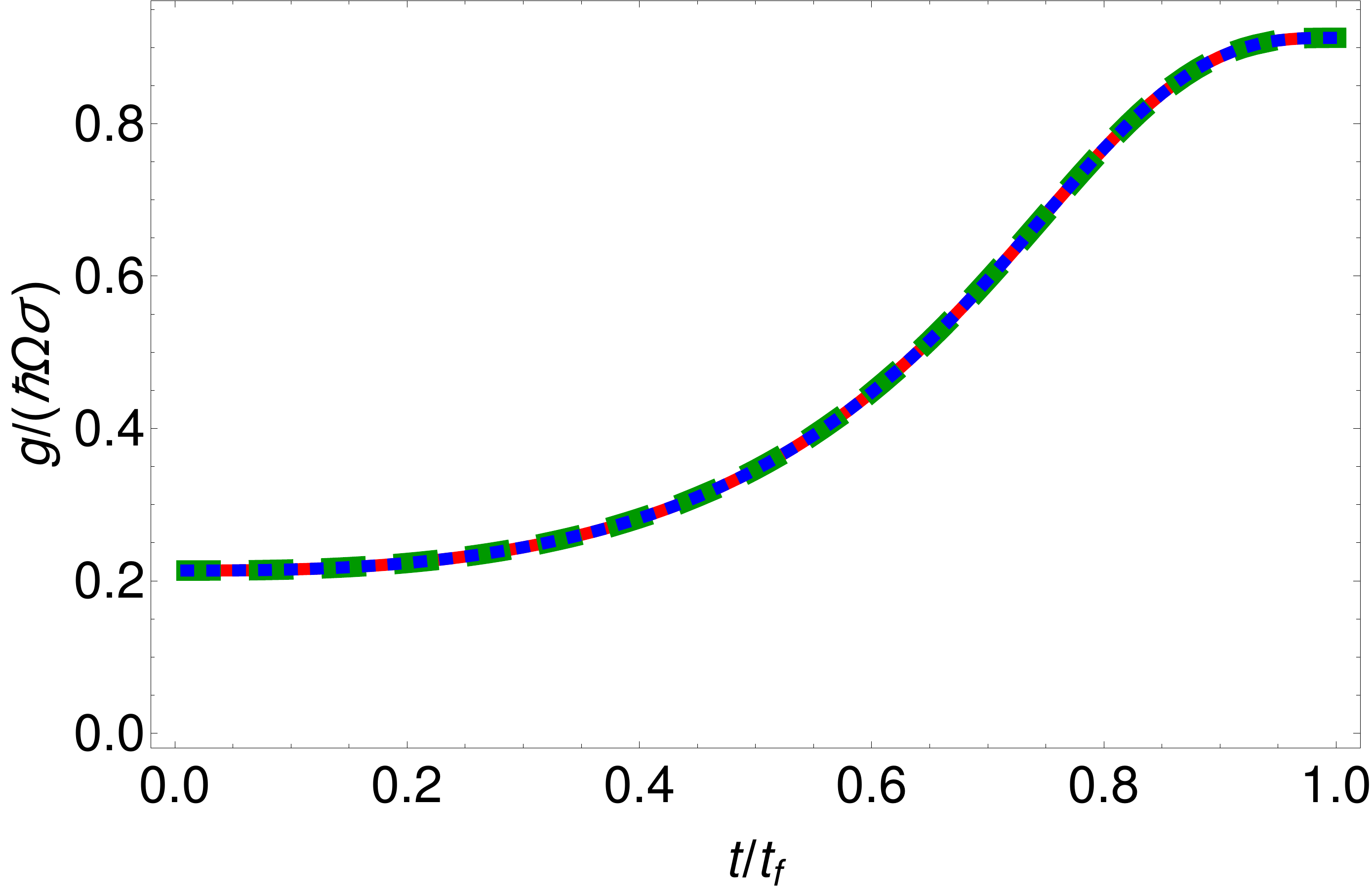}
	\end{center}
\caption[Control functions for loading particles into trap.]{Loading particles into an external trap: (a) $\omega(t)$  versus $t/t_{f}$; (b) $g(t)$ versus $t/t_{f}$. Final time: $t_{f} = 0.55$ $T$ (red solid line), $t_{f} = 1.10$ $T$ (green dashed line), $t_{f} = 2.19$ $T$ (blue dotted line).}
\label{fig:dynam_fns_loading}
\end{figure}
The goal here is to load the particle or condensate from a lattice into an external harmonic trap without final motional excitation. We start with the external harmonic trap having a frequency of $\omega(0)=0$ at initial time $t=0$ and $\omega(t_{f})=\omega_{f}$ at final time $t_{f}$. The position of the trap remains unchanged in a well of the lattice $q_{0}(t) = 0$. Now considering Eqs. (\ref{eq:Aux_eqns_1}) and (\ref{eq:Aux_eqns_2}) we see the auxiliary function $q_{c}(t)$ can be set $q_{c}(t) = q_{0}(0) = 0$. This leaves us with Eq. (\ref{eq:Aux_eqns_1}); and so $\rho(t)$ must satisfy the following boundary conditions,
\begin{eqnarray}
\rho(0) &=& 1  ;\\
\rho(t_{f}) &=& \sqrt{\dfrac{\tilde{\omega}(0)}{\tilde{\omega}(t_{f})}}; \\
\dot{\rho}(0) &=& \dot{\rho}(t_{f}) = 0 ;\\
\ddot{\rho}(0) &=&\ddot{\rho}(t_{f}) = 0.
\label{eq:bound_conds_2}
\end{eqnarray}
We choose a polynomial of minimal degree that satisfies the above boundary conditions for $\rho(t)$. We can then substitute this $\rho(t)$ to find the virtual frequency $\tilde{\omega}(t)$ as a function of time according to Eq. (\ref{eq:w_control}). This approach corresponds to having to tune the external trap frequency according to 
\begin{equation}
\omega(t)^{2} = \tilde{\omega}(t)^{2} + \Omega^{2},
\end{equation}
We call this approach the shortcut scheme. Note that if the harmonic approximation is exact, then the corresponding shortcut scheme will achieve a fidelity of $F = 1$ in arbitrarily short timescales. For the case of the Gross-Pitaevskii equation we must also tune the atom-atom interaction according to Eq. (\ref{eq:Aux_eqns_3}). We set the parameters as follows; the lattice height $U_{0}/(\hbar \Omega) = 547.7$, the final frequency of the external harmonic trap is chosen as $\omega_{f} = 18.257 \Omega$.  We now simulate the full Schr\"odinger and Gross-Pitaevskii equations using exact initial states obtained by numerically solving the relevant stationary equations.
\begin{figure}[!t]
	\begin{center}
	(a)\\
		\includegraphics[width = 0.8 \linewidth]{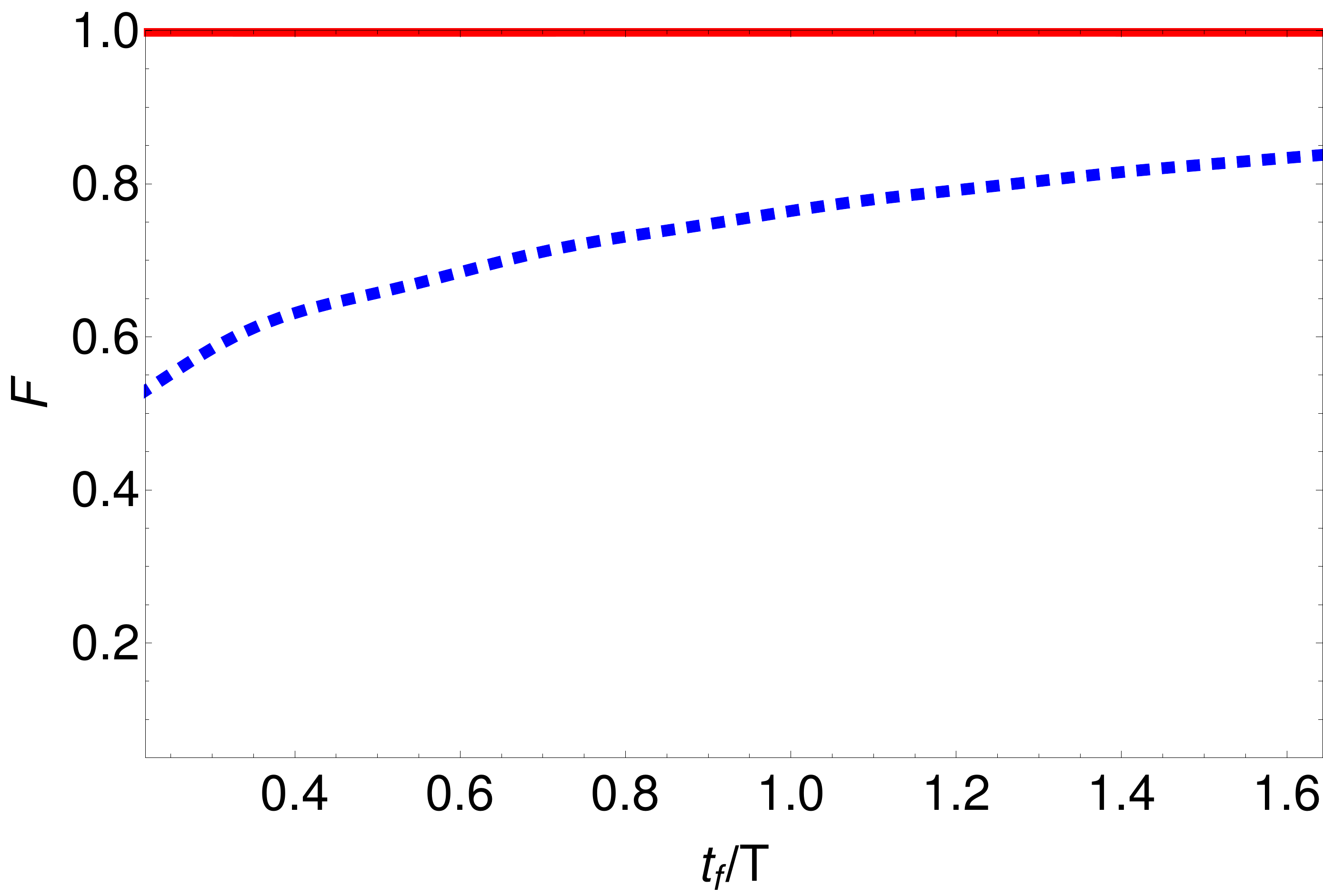}\\
	(b)\\
		\includegraphics[width = 0.8 \linewidth]{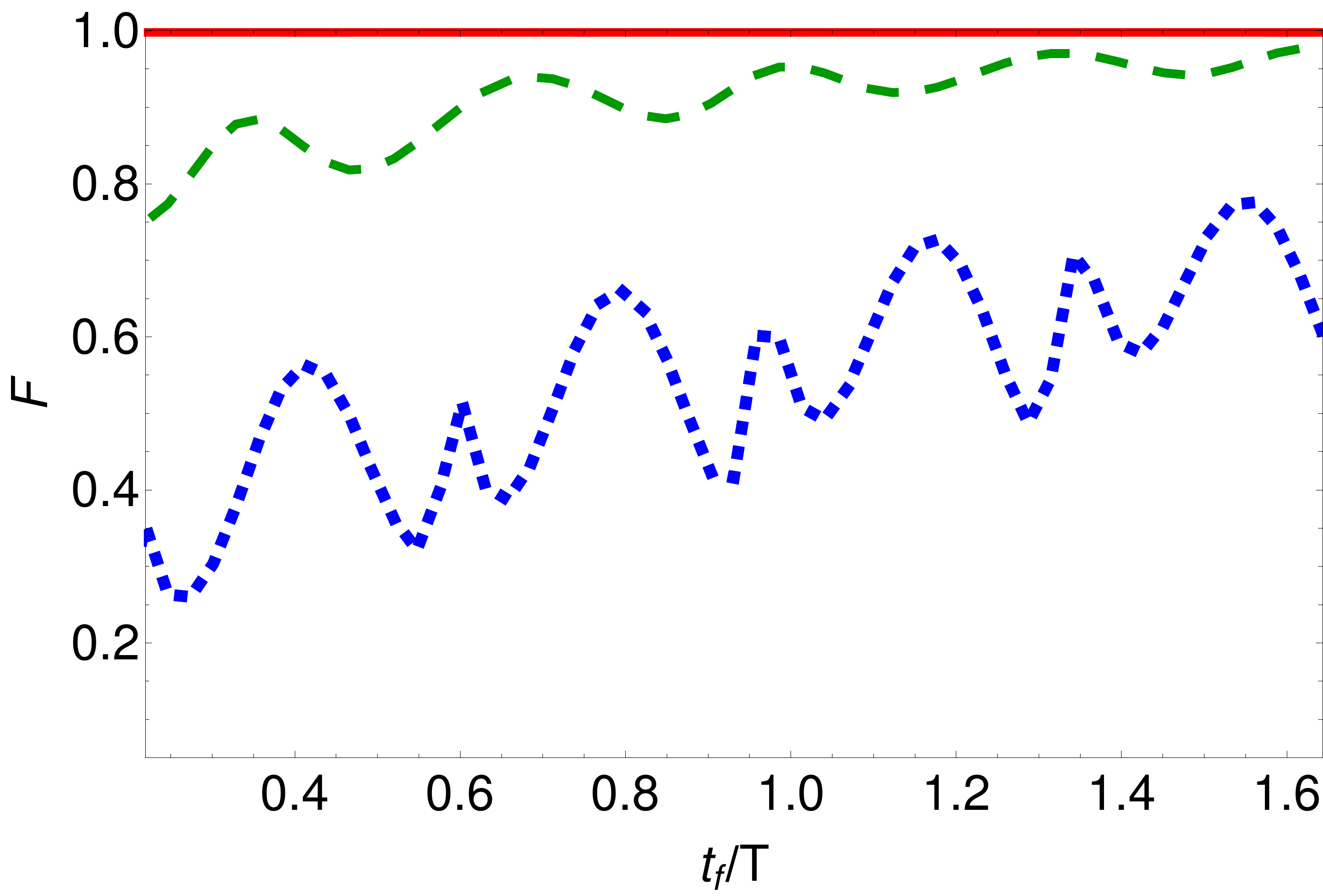}
	\end{center}
\caption[Compression fidelity $F$ versus final time $t_{f}$]{Loading particles into an external trap: Fidelity $F$ versus final time $t_{f}$; (a) $g(t) = 0$; (b) $g_{f} = 0.91$ $\hbar \Omega \sigma$. The shortcut scheme (red solid line), the adiabatic scheme (blue dotted line), the constant $g$ scheme (green dashed line).}
\label{fig:comp_fid}
\end{figure} 
In Fig. \ref{fig:dynam_fns_loading} we plot the control functions $\omega(t)$ and $g(t)$ for different values of $t_{f}$ and we see that the $\omega(t)$ function changes for different values of $t_{f}$ but the $g(t)$ function remains the same; this is because the auxiliary function $\rho(t)$ is a polynomial of $t/t_{f}$. In Fig. \ref{fig:comp_fid} we plot the fidelity $F$ as a function of final time $t_{f}$, in (a) for $g=0.0$ and in $(b)$ for $g_{f} = 0.913 (\hbar \Omega \sigma)$.
For comparison we also consider two alternate schemes, first varying the trap frequency $\omega(t)$ adiabatically according to
\begin{align*}
\omega(t) = (\omega(t_{f}) - \omega(0))\sin\left(\dfrac{t \pi}{2 t_{f}}\right)^{2} + \omega(0),
\end{align*} 
and second varying the $\omega(t)$ as in the shortcut protocol but with constant atom-atom interaction $g(t) = g_{f} = 0.91 \hbar \Omega \sigma$. We see in Fig. \ref{fig:comp_fid} that the shortcut scheme performs well achieving fidelities of $F \geq 0.99$ for all times. This is to be expected as the harmonic approximation in this case is very good. The adiabatic scheme however performs poorly in comparison; in the $g=0$, case see Fig. \ref{fig:comp_fid} (a) it achieves fidelities $F < 0.83$ for all time-scales shown. In Fig. \ref{fig:comp_fid} (b) we see that the third scheme of varying the $\omega(t)$ according to the shortcut protocol but with $g(t) = g_{f}$ constant doesn't achieve the same high fidelities as the full shortcut protocol but still performs better than the adiabatic case. In the case of $g = g_{f} = 0.91$ $\hbar \Omega \sigma$, we see that both the adiabatic and the constant $g$ approach are oscillatory. When considering the atom-atom interaction in Fig. \ref{fig:comp_fid} (b), we see that the adiabatic scheme performs worse than in the $g=0$ case; the shortcut scheme however still achieves the high fidelities on all time-scales. 
\subsection{Shifting across Lattice Site}
In this subsection we want to shift the atom or condensate from one lattice site to its nearest neighbour. This procedure could be concatenated to achieve transport of the atom or condensate over a number of lattice sites. The external harmonic trap will thus start at $q_{0}(0) = 0$ and at final time will be at $q_{0}(t_{f}) = \pi \sigma$. In addition the frequency of the external harmonic trap should be the same at initial and final time $\omega_{0} = \omega(t_{f}) = \omega(0)$.
\begin{figure}[!t]
	\begin{center}
		(a)\\
		\includegraphics[width = 0.8 \linewidth]{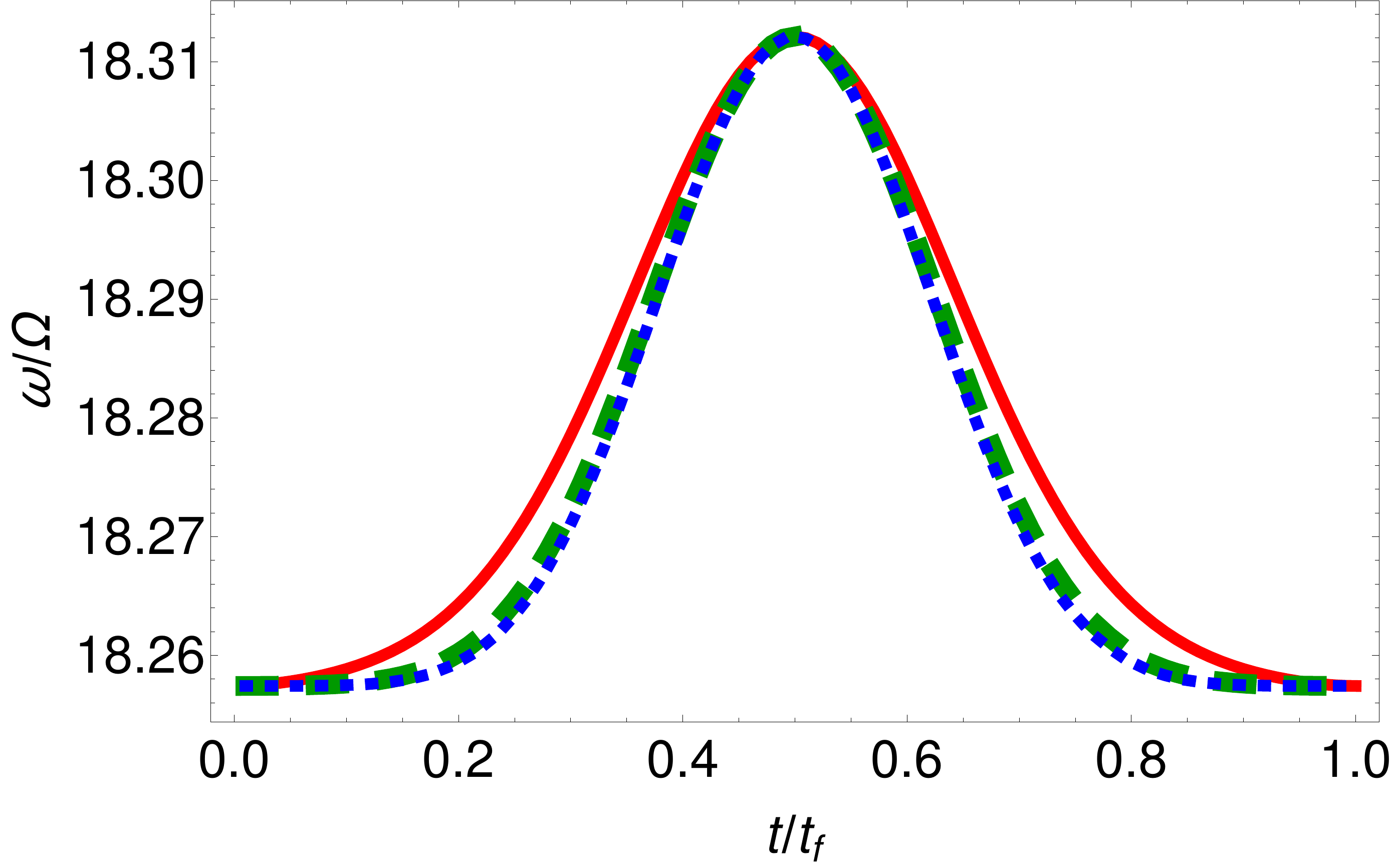}\\
		(b)\\
		\includegraphics[width = 0.8 \linewidth]{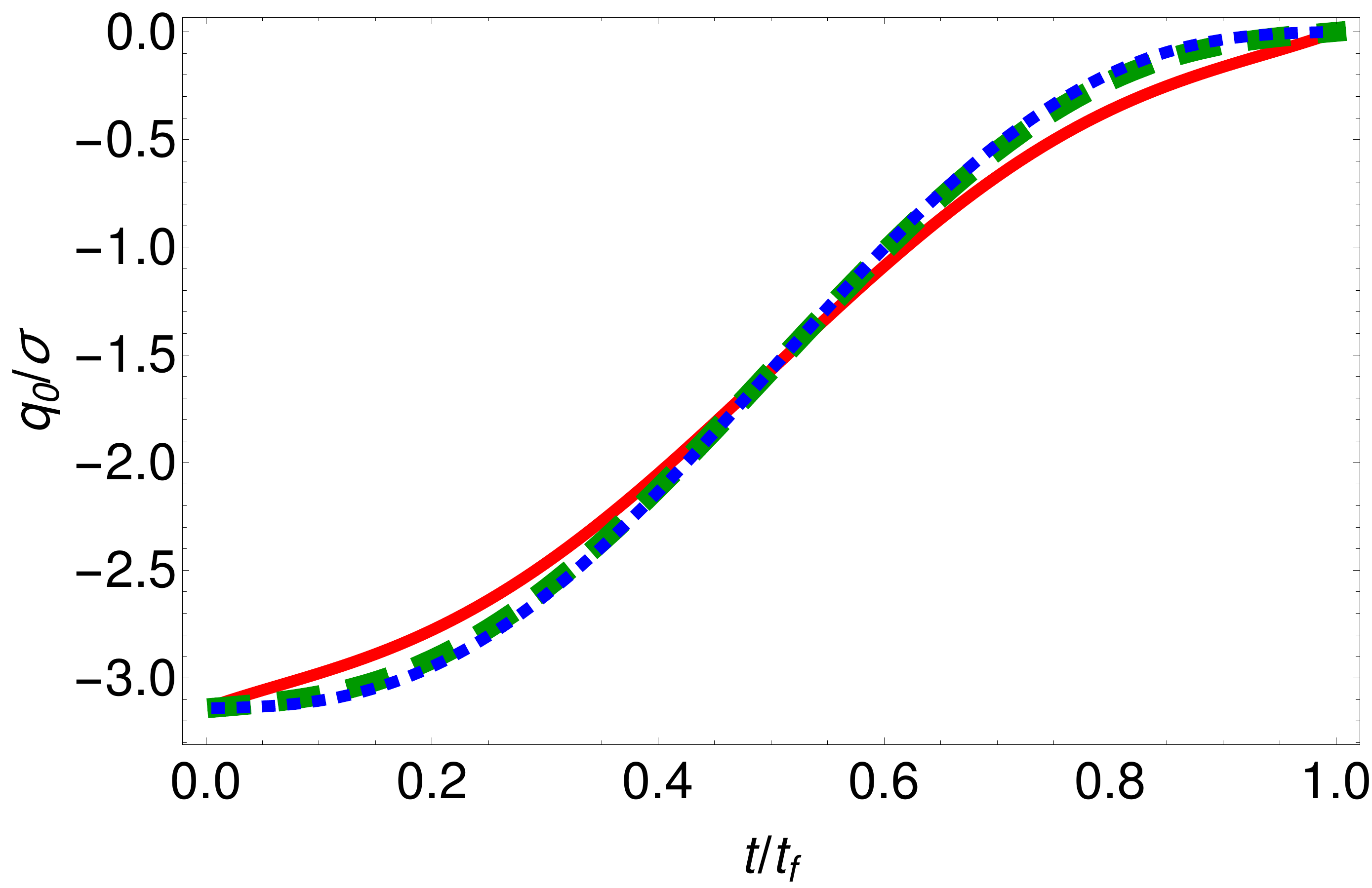}
	\end{center}
\caption[Control functions for transporting particles across lattice site.]{Shifting particles across a lattice site: (a) $\omega(t) \text{ versus }t/t_{f}$; (b) $q_{0}(t) \text{ versus } t/t_{f}$. Final time: $t_{f} = 0.55$ $T$(red solid line), $t_{f} = 1.10$ $T$ (green dashed line) $t_{f} = 2.19$ $T$ (blue dotted line).}
\label{fig:dynam_fns_transport}
\end{figure}
We design a shortcut scheme we call the variable frequency scheme, that solves the Eqs. (\ref{eq:Aux_eqns_1}) and (\ref{eq:Aux_eqns_2}) exactly. The key idea here is to alter the harmonic trap frequency $\omega(t)$ in such a way that the virtual trap frequency stays constant \textit{i.e.} $\tilde{\omega}(t) = \tilde{\omega}(0) = \omega(0)^{2} + \Omega^{2}$. This allows us to solve Eq. (\ref{eq:Aux_eqns_1}) by setting $\rho = 1$. In the case of a condensate, this has the added benefit that there is no need to tune the atom-atom interaction in time, as $g(t) = g(0)/\rho(t) = g$. The boundary conditions on the auxiliary function $q_{c}(t)$ are 
\begin{align*}
&q_{c}(0) = 0; \hspace{3 cm} q_{c}(t_{f}) = \pi \sigma;\\
&\dot{q}_{c}(0) = 0; \hspace{3 cm} \dot{q}_{c}(t_{f}) = 0;\\
&\ddot{q}_{c}(0) = 0; \hspace{3 cm} \ddot{q}_{c}(t_{f}) = 0.\\
\end{align*}
We choose a polynomial solution of minimal degree to fulfil these boundary conditions and so we can calculate the position virtual trap centre as
\[
x_{\text{min}}(t) = q_{c}(t) + \dfrac{1}{\tilde{\omega}^{2}} \ddot{q}_{c}(t) 
\]
and the control function for the actual trap centre is then given by
\begin{eqnarray}
\lefteqn{q_{0}(t) = x_{min}(t)} & & \nonumber\\
& & + \dfrac{\Omega^{2}}{ \omega(t)^{2}}  \sin\left(\dfrac{x_{min}(t)}{\sigma}\right)
\cos\left(\dfrac{x_{min}(t)}{\sigma}\right).
\label{eq:q0eqn}
\end{eqnarray}
In addition, for this approach we vary the trap frequency according to
\begin{eqnarray}
\omega(t)^{2} &=& \tilde{\omega}^{2} + \Omega^{2} \cos\left(\dfrac{2 x_{min}(t)}{\sigma}\right)\nonumber\\
&=& \omega_{0}^{2} + \Omega^{2}\left[1 + \cos \left(\dfrac{2 x_{m}(t)}{\sigma}\right)\right].
\label{eq:omegaeqn}
\end{eqnarray}
Both the trap centre position $q_{0}(t)$ and trap frequency $\omega(t)$ control functions are shown in Fig. \ref{fig:dynam_fns_transport} with the frequency $\omega(t)$ shown in (a) and the trap centre position $q_{0}(t)$ shown in (b).
\subsubsection{Fidelities based on full Schr\"odinger/G-P equation}
We now simulate the full Schr\"odinger and Gross-Pitaevskii equations with an exact initial and final state using these schemes i.e. we assume first that the previous loading of the particles into the trap had fidelity one.\\
\begin{figure}[!t]
	\begin{center}
		(a)\\
		\includegraphics[width = 0.8 \linewidth]{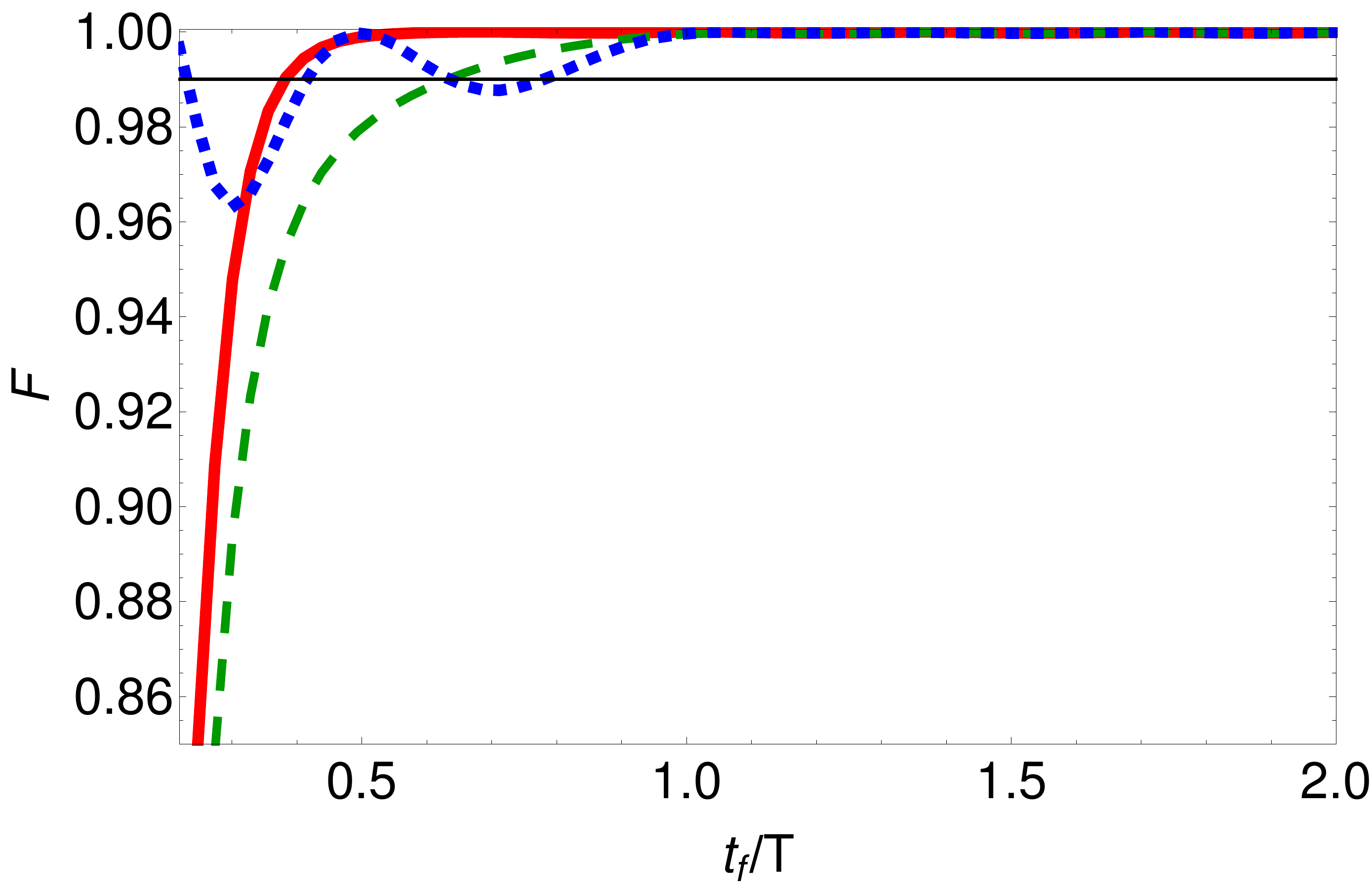}\\
		(b)\\
		\includegraphics[width = 0.8 \linewidth]{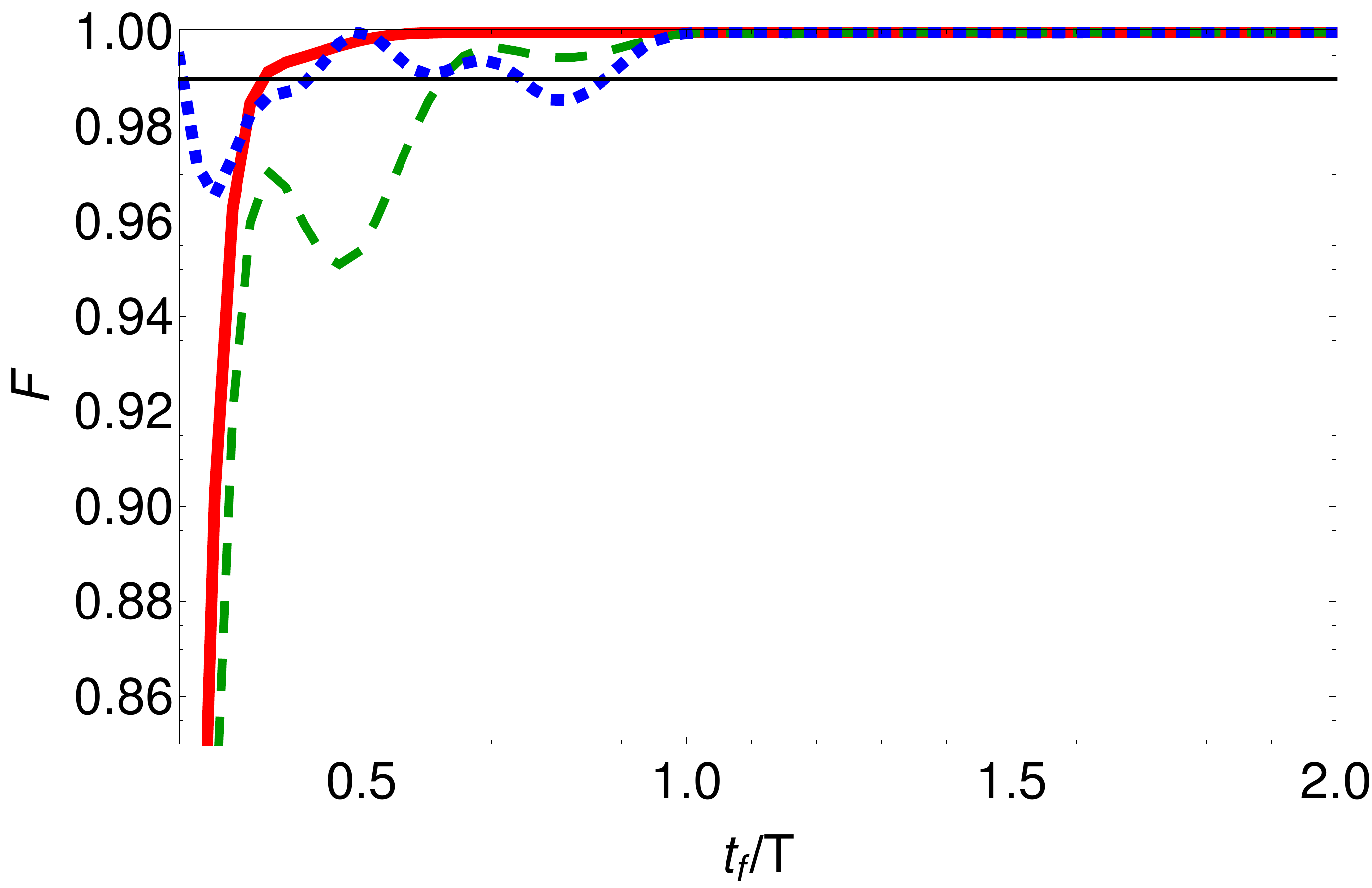}\\
	\end{center}
\caption[Transport fidelity $F$ versus final time $t_{f}$.]{Shifting particles across a lattice site: Fidelity $F$ versus final time $t_{f}$, (a) $g = 0$; (b) $g = 0.91$ $\hbar \Omega \sigma$. The shortcut scheme (red solid line), the first (blue dotted line) and second (green dashed line) constant frequency approximations.}
\label{fig:trans_fid}
\end{figure}
The fidelities for different final times $t_{f}$ are shown in Fig. \ref{fig:trans_fid} (a) ($g=0$) and (b) ($g=0.91 (\hbar \Omega \sigma))$ as the red solid line. This variable frequency scheme approach performs very well achieving high fidelities even on very short time-scales. For a more in-depth look, we examine the threshold time $t_{0.99}$ which is defined as the time for which the fidelity $F \geq 0.99$ for all times $t \geq t_{0.99}$. We plot this quantity $t_{0.99}$ for different frequencies $\omega_{0}$ in Fig. \ref{fig:threshold_fid}. We see that the threshold time $t_{0.99}$ decreases as initial trapping frequency $\omega_{0}$ is increased for both values of $g$ shown.
\subsubsection{Approximated transport schemes}
We now also consider two approximated transport schemes, assuming $\omega_{0} \gg \Omega$; in particular we look at these because these two schemes do not require the tuning of the external harmonic frequency $\omega(t)$ during the transport. \\
The first approximated scheme is achieved by neglecting the $\Omega^{2}$ term in Eq. (\ref{eq:omegaeqn}), as $\omega_{0} \gg \Omega$ leading to $\omega(t) = \omega(0)$. This means that we are implementing the same $q_{0,A1}(t) = q_{0}(t)$ function as the variable frequency scheme (shown in Fig. \ref{fig:dynam_fns_transport} (b)), but still keeping the frequency $\omega(t)=\omega(0)$ constant during transport. This we label "first constant frequency approximation".\\
The second approximation is similar, we also fix $\omega(t) = \omega(0)$ but here we also neglect any terms proportional to $\Omega^{2}/\omega_{0}^{2}$ in Eq. (\ref{eq:q0eqn}), giving us the following trap centre function
\begin{equation}
q_{0,A2}(t) = x_{min}(t) = q_{c}(t) + \dfrac{1}{\omega_{0}^{2}} \ddot{q}_{c,0}(t).
\end{equation} 
We call this the "second constant frequency approximation".\\
\begin{figure}[!t]
	\begin{center}
		\includegraphics[width = 0.8 \linewidth]{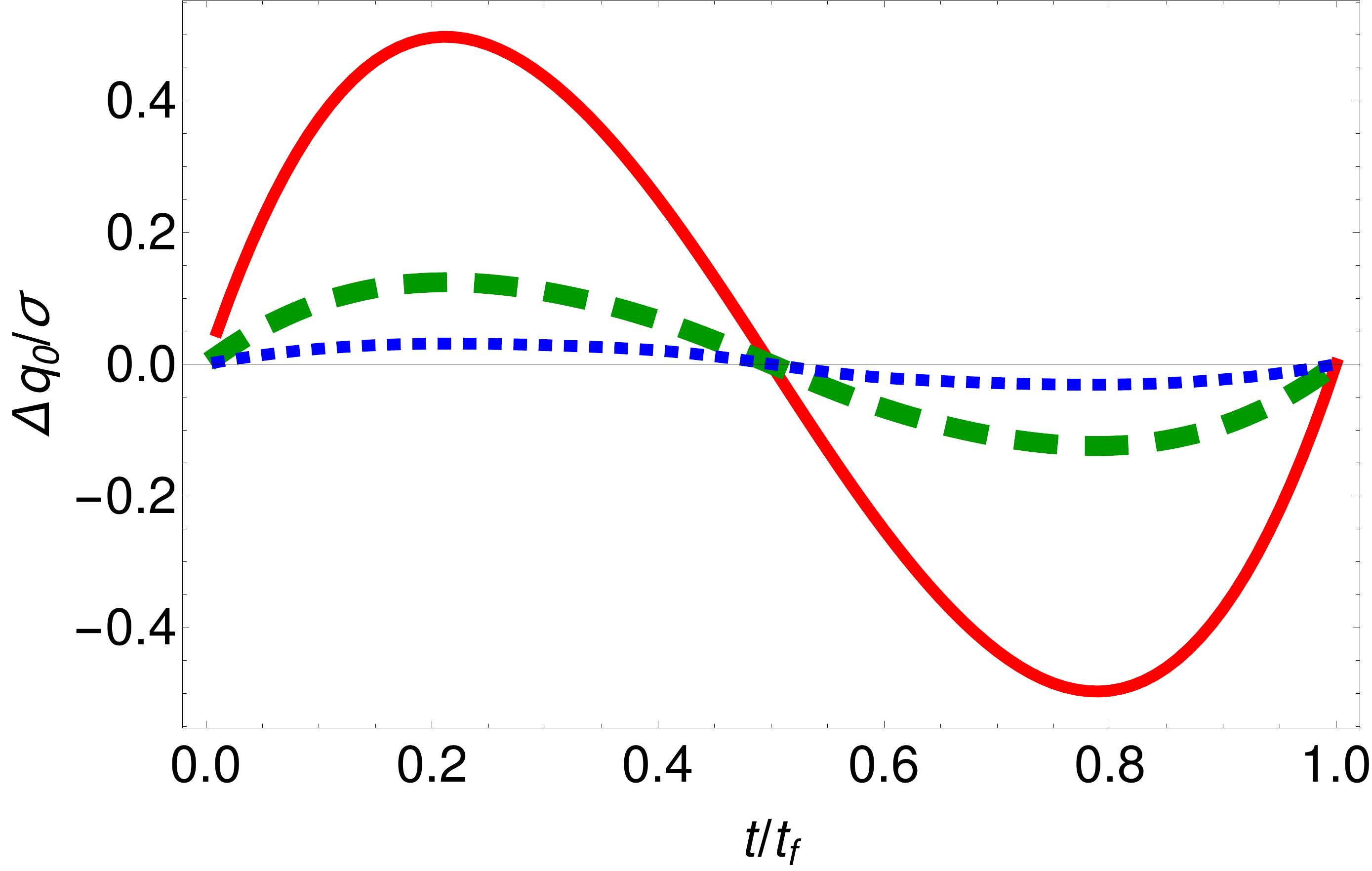}
	\end{center}
\caption[Difference in $q_{0}(t)$ between different transport protocols.]{Shifting particles across a lattice site: Difference between variable frequency scheme and "second constant frequency approximation":  $\Delta q_{0}(t)$ versus $t/t_{f}$. Final time: $t_{f} = 0.55$ $T$(red solid line), $t_{f} = 1.10$ $T$ (green dashed line) $t_{f} = 2.19$ $T$ (blue dotted line).}
\label{fig:exp_cont}
\end{figure}
The particular strength of the above two approximations is that there is no longer any need to control the trap frequency $\omega$ or the atom-atom interaction $g(t)$ during the transport. Instead the only varying function is the trap centre position $q_{0}(t)$. Both schemes will result in different trap trajectories, the "first constant frequency approximation" will have the same trap trajectory $q_{0,A1}(t)$ as the variable frequency scheme derived earlier and shown in Fig. \ref{fig:dynam_fns_transport}(b). However the "second constant frequency approximation"  is different. The difference between the two trajectories $q_{0,A1}$ and $q_{0,A2}$ is seen in Fig. \ref{fig:exp_cont} for different final times $t_{f}$. We see that with increasing final time $t_{f}$, the differences between the two schemes decrease.\\
\\
Similarly to the previous subsection, we now solve the exact Schr\"odinger and Gross-Pitaevskii equations numerically, with exact initial states. We have plotted the fidelities in Fig. \ref{fig:trans_fid} for both the two approximation schemes together with the variable frequency scheme. Both approximation schemes result in high fidelities for both $g$ values shown but perform significantly worse than the variable frequency scheme described earlier. They both achieve fidelities of $F \geq 0.99$ but on longer time-scales than the variable frequency scheme.\\
\\
\begin{figure}[!t]
	\begin{center}
	(a)\\
		\includegraphics[width = 0.8 \linewidth]{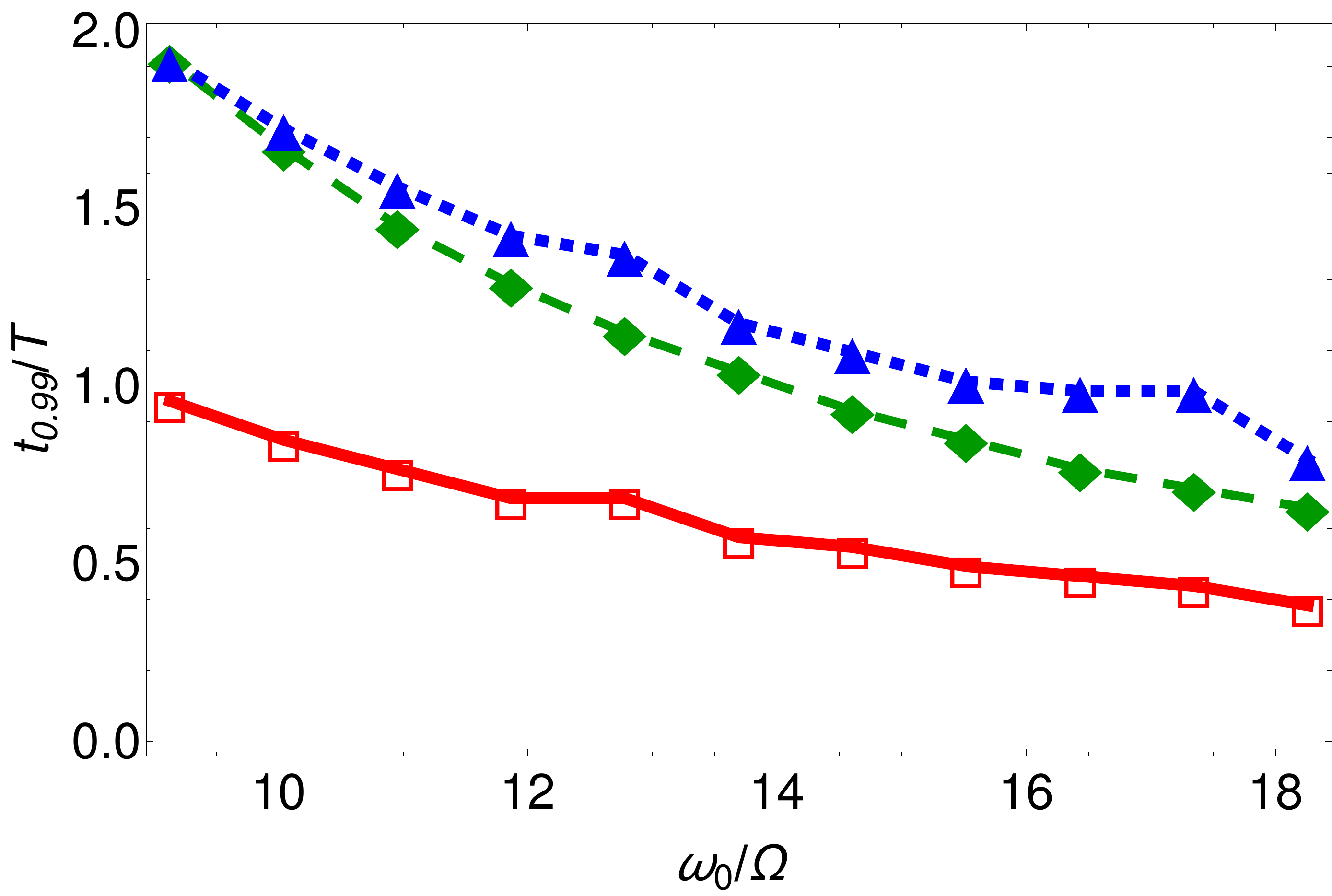}\\
		(b)\\
		\includegraphics[width = 0.8 \linewidth]{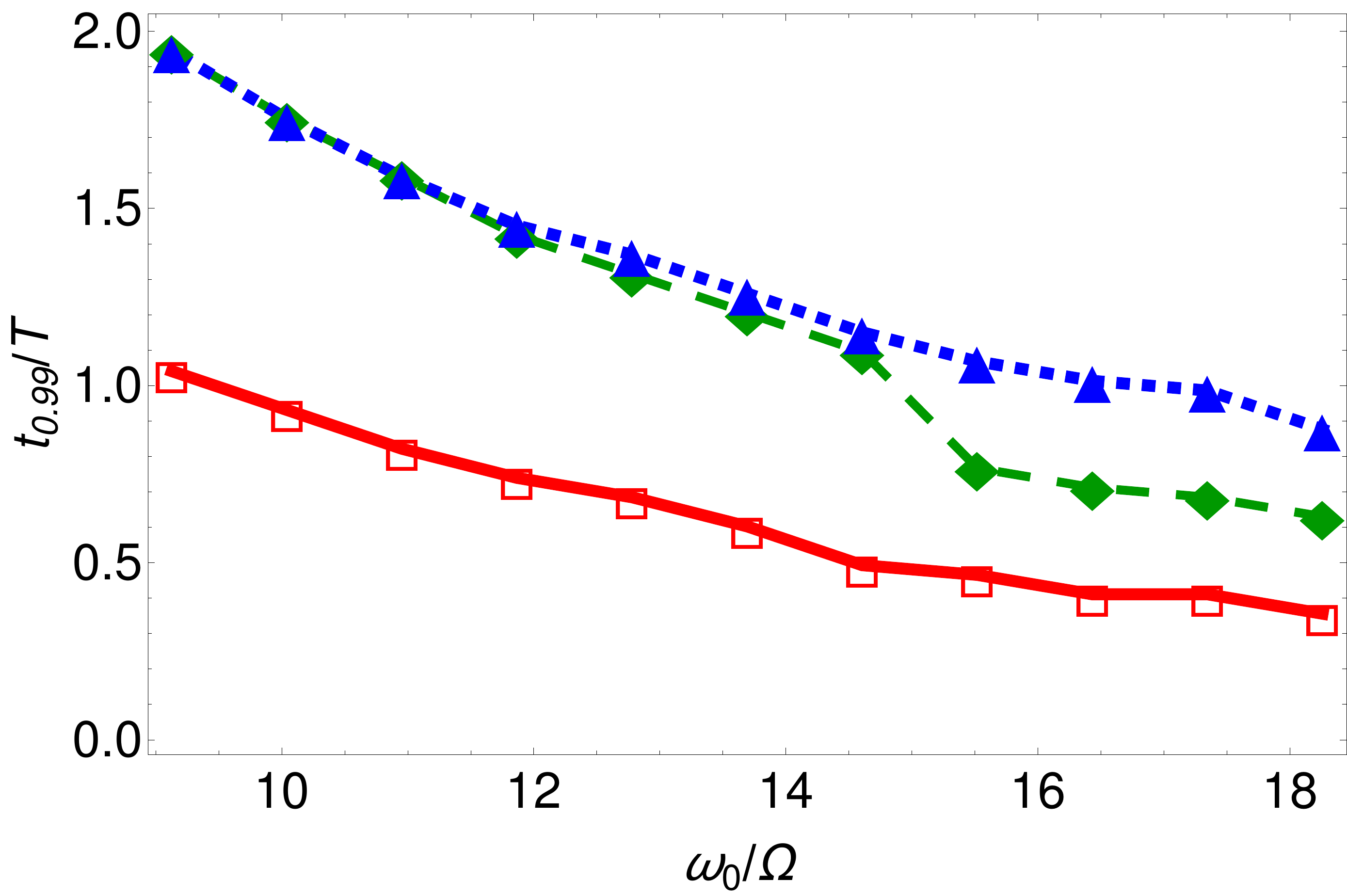}\\
	\end{center}
\caption[Transport threshold time $t_{0.99}$ versus $\omega_{0}$.]{Shifting particles across a lattice site: Threshold time $t_{0.99}$ versus $\omega_{0}$ (a) $g=0$; (b) $g = 0.91$ $\hbar \Omega \sigma$. Exact scheme (red boxes connected with a solid line), "first constant frequency approximation" (green diamonds connected by a dashed line), "second constant frequency approximation" (blue triangles connected by a dotted line).}
\label{fig:threshold_fid}
\end{figure}
We again examine the threshold time $t_{0.99}$ in Fig. \ref{fig:threshold_fid}. We see that while the variable frequency scheme performs the best, the two approximate schemes still give a threshold time $t_{0.99}$ slightly larger than the variable frequency scheme and do not require control of trap frequency $\omega(t)$. This may prove useful in situations where the frequency of the trapping potential is difficult to tune. It appears in Fig. \ref{fig:threshold_fid} (a) and (b) that the "first constant frequency approximation" performs at least as well as the "second constant frequency approximation" and in some circumstances such as lower $g$ and higher $\omega$, it performs better. By comparison using an adiabatic scheme where the trap centre is varied according to 
\begin{align*}
q_{0}(t) = (q_{0}(t_{f}) - q_{0}(0)) \sin\left(\dfrac{t \pi }{2 t_{f}}\right)^{2} + q_{0}(0)
\end{align*}
the threshold time $t_{0.99}$ is much higher, the scheme achieves the threshold fidelity around $t_{0.99} \approx 10.4$ $T$.
\subsubsection{Robustness}
\begin{figure}[!t]
	\begin{center}
	(a)\\
		\includegraphics[width = 0.8 \linewidth]{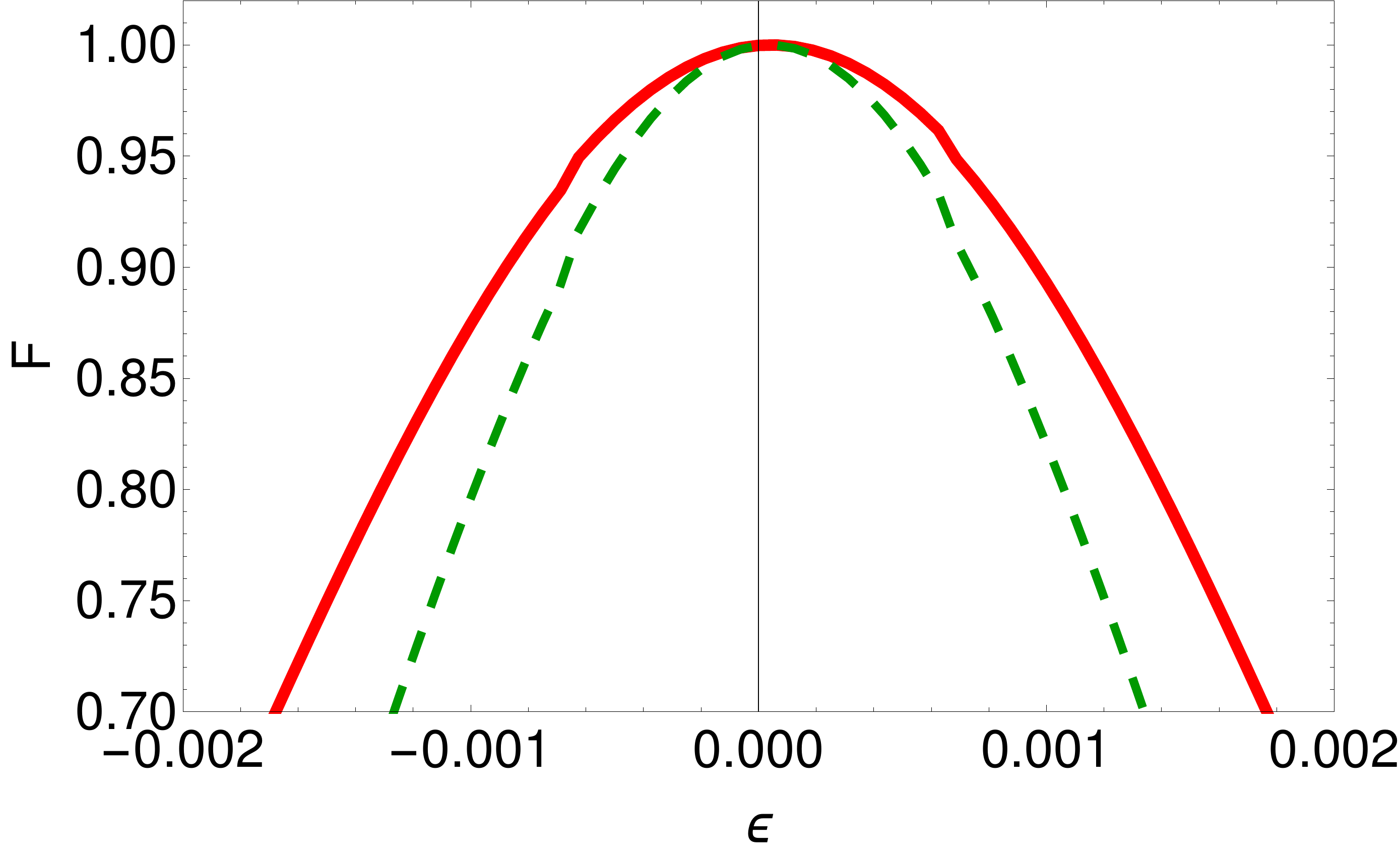}\\
		(b)\\
		\includegraphics[width = 0.8 \linewidth]{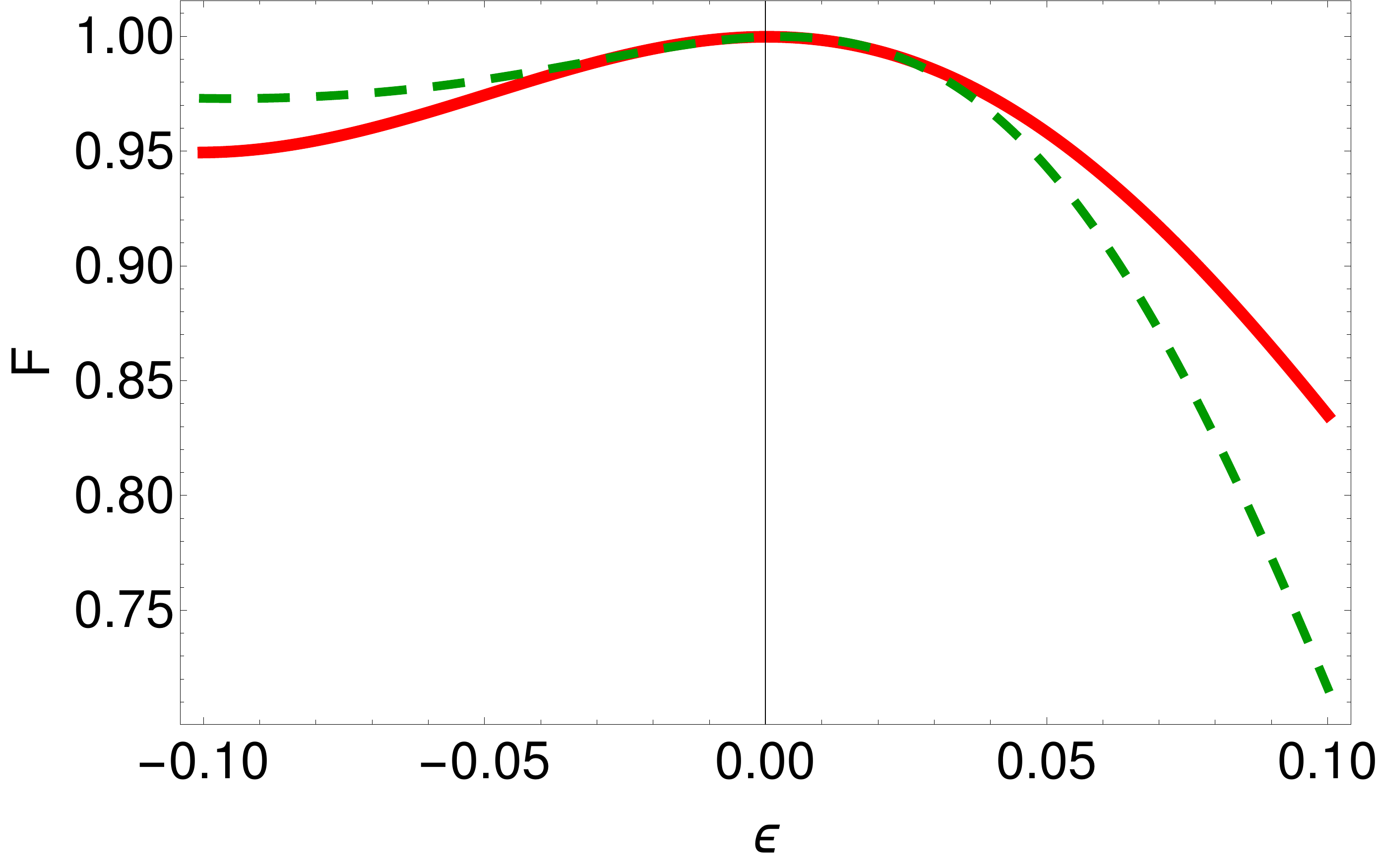}
	\end{center}
\caption[Transport fidelity $F$ versus perturbation $\epsilon d , \epsilon$. ]{ Moving particles across a lattice site: Fidelity $F$ versus perturbation $\epsilon$ (a) error in $q_{0}(t)$ (b) error in $\omega(t)$. $g = 0$ (red solid line), $g = 0.91$ $\hbar \Omega \sigma$ (green dashed line).}
\label{fig:robustness_fid}
\end{figure}

In this subsection we examine the robustness of the variable frequency scheme for shifting the trap. We will consider an error in the position $q_{0}$ and later in the frequency $\omega$ during the transport. First let us consider an error in the trap position $q_{0}$ of the form
\begin{equation}
q_{0}(t) = q_{0,\text{exact}}(t) + d \epsilon, \hspace{1cm} 0 < t < t_{f}
\end{equation}
where $\epsilon$ is a small perturbation parameter and $d = \sigma \pi$ is the distance between the two lattice sites. The perturbation only acts during the transportation, at boundary times $q_{0}(0) = q_{0,\text{exact}}(0)$ and $q_{0}(t_{f}) = q_{0,\text{exact}}(t_{f})$. The  frequency of the external harmonic trap is chosen as $\omega_{0} = 18.257$ $\Omega$ and the final time is $t_{f} = 1.10$ $T$. We have plotted the fidelity $F$ versus the perturbation $\epsilon$ in Fig. \ref{fig:robustness_fid}(a). We see that the the region close to $\epsilon = 0$ retains high fidelities as expected showing this variable frequency protocol is stable against this perturbation. \\
\\
As a second form of perturbation let us consider an error in the trap frequency $\omega$ of the form
\begin{equation}
\omega(t) = \omega_{exact}(t) (1 + \epsilon), \hspace{1cm} 0 < t < t_{f}.
\end{equation}
Here $\epsilon$ is a small perturbation parameter that changes the frequency of the external trap. Again the system is perturbed only during the shifting. We have plotted the fidelity $F$ versus the perturbation $ \epsilon$ in Fig. \ref{fig:robustness_fid} (b). There is an asymmetry in the fidelity in both the case of $g = 0$ and $g = 0.91$  $\hbar \Omega \sigma$. The scheme for $g = 0.91$ $\hbar \Omega \sigma$ performs better than the scheme for $g = 0.0$ for perturbations with $\epsilon < 0$, but for perturbations with $\epsilon > 0$ the $g=0.0$ scheme achieves higher fidelities. In summary the proposed variable frequency scheme is robust and stable against perturbations in the external trap trajectory and frequency
\subsection{Unloading onto lattice}
In this section we now attempt to open the external harmonic trap to unload the particles back onto the lattice after transport. We start with the frequency of the harmonic trap $\omega(0) = \omega_{0} > 0$ at initial time $t=0$ and finish with $\omega(t_{f}) = 0$ at final time $t_{f}$. The position of the external trap stays constant in the well of a lattice such that $q_{0}(t) = n \pi \sigma$, $n \in \mathcal{N}$ for all $t \geq 0$. \\
There is no change in position of the trap so the auxiliary function $q_{c}(t)$ can be chosen to be constant $q_{c}(t) = q_{0}(0)$. We can then pick the auxiliary function $\rho(t)$ to satisfy the following boundary conditions
\begin{eqnarray}
\rho(0) = 1  ;\hspace{1 cm} \rho(t_{f}) = \sqrt{\dfrac{\tilde{\omega}(0)}{\tilde{\omega}(t_{f})}}; \nonumber\\
\dot{\rho}(0) = \dot{\rho}(t_{f}) = 0 ; \hspace{1 cm}\ddot{\rho}(0) =\ddot{\rho}(t_{f}) = 0.\nonumber
\end{eqnarray}
Again we choose a polynomial $\rho(t)$ of minimal degree to fulfil these boundary conditions. This approach corresponds to tuning the external harmonic trap frequency as follows
\begin{equation*}
\omega(t) = \tilde{\omega}(t) + \Omega^{2}.
\end{equation*}
In the case of the atom-atom interaction, we tune $g(t)$ according to $g(t) = g_{0}/\rho(t)$ following from earlier results.
The unloading is a direct reverse of the previous loading and in the sense that each of the auxiliary functions $\omega(t)$ and $g(t)$ is the time reversed function from the subsection \ref{subsec:loading}. Again, if the harmonic approximation is exact, the fidelity of the scheme would be $F = 1$, independent of final time $t_{f}$. The $g=0$ case is the time reversal of the loading and so the fidelity is the same as in Fig. \ref{fig:comp_fid}(a).
We now, as in previous sections, simulate the full dynamics of the system using the Schr\"odinger and Gross-Piteavskii equations with an exact initial state for $g = 0.91$ $\hbar \Omega \sigma$. The initial frequency of the harmonic trap is chosen to be $\omega_{i} = 18.257 \Omega$.
\begin{figure}[!t]%[t]
	\begin{center}
		\includegraphics[width = 0.8 \linewidth]{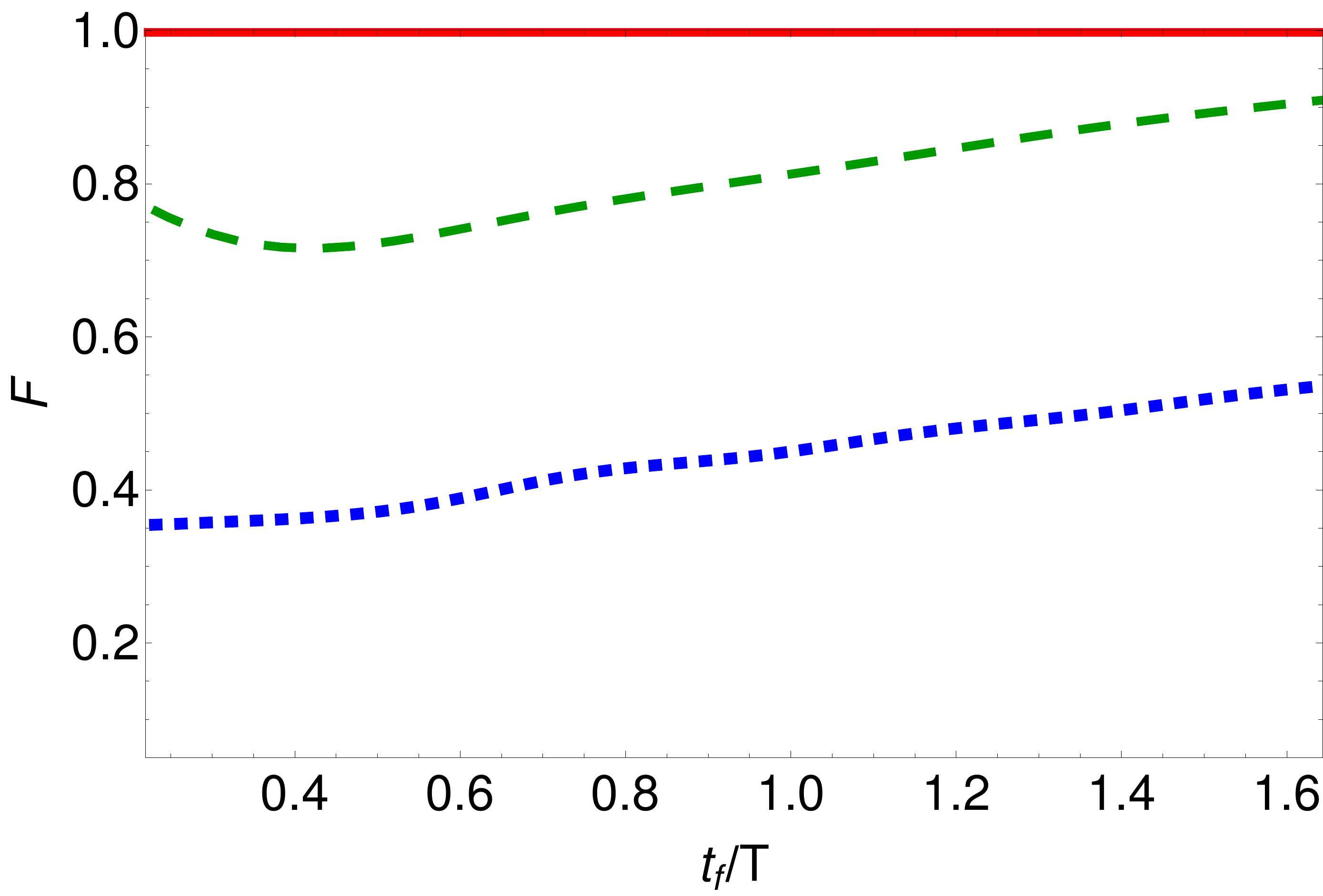}
	\end{center}
\caption[Unloading fidelity $F$ versus final time $t_{f}$.]{Unloading particles back onto lattice: Fidelity $F$ versus final time $t_{f}$, $g_{0} = 0.91$ $\hbar \Omega \sigma$. The shortcut scheme (red solid line), the adiabatic scheme (blue dotted line), the constant $g$ scheme (green dashed line).}
\label{fig:exp_fid}
\end{figure}
We have plotted the fidelity $F$ versus final time $t$ in Fig. \ref{fig:exp_fid}. Similarly to the earlier case of loading particles into the trap, the shortcut scheme achieves a stable fidelity of $F \geq 0.99$ for all times shown. However, in the case of unloading particles back onto the lattice, the adiabatic scheme is more stable. In the earlier figure, Fig. \ref{fig:comp_fid}, for loading we saw that, for $g = 0.91$ $\hbar \Omega \sigma$, the fidelity as a function of time varies more and does not display the almost monotonic behaviour seen in Fig. \ref{fig:exp_fid}. However, in both the loading and unloading, the adiabatic protocol doesn't perform well when compared with the shortcut protocol or the constant $g$ protocol. 

\section{Conclusion}
We have proposed a method utilizing STA for the fast and robust transport of atoms or for a Bose-Einstein condensate across a optical lattice by using an external trapping potential. To do this we have broken the transport process into three independent building blocks: first loading a particle from a lattice site into an external trapping potential, then shifting the particle across the lattice and finally unloading the particle from the external trapping potential back on to a lattice site. We then applied methods from STA to each of these building blocks to derive approximated control schemes for the external trap. Concatenating all three of the different building blocks we can transport particles from one lattice site, trap them and then shift them to another and finally unload them into the target lattice site. Alternative schemes to achieve similar fidelities but requiring less control were also considered. The sensitivity of the protocols with respect to trap centre control and trapping frequency were investigated and the protocols were shown to be robust against these errors. In future work we will optimise the stability of the transport across the lattice versus noise following the formalism in \cite{noise,lattice_noise}.
\section*{Acknowledgements}
We would like to thank David Rea for reading and comments on the manuscript. TD acknowledges the support of the Irish Research Council (GOIPG/2015/3195).
\appendix
\begin{widetext}
\section{STA applied to Gross-Piteavskii Equation}
\label{BEC_extension}
In the following we will review how STA techniques can be applied to a Bose-Einstein condensate; this is based on combining the results of previous work transporting a condensate \cite{bec_transport} and varying the trap parameters for a condensate \cite{bec_trap_var}. We make the same harmonic approximation as in Eq. (\ref{eq:harm_approx}) so that the wavefunction evolves according to
\begin{equation}
i \hbar \partial_{t} \psi(x,t) = \left(- \dfrac{\hbar^{2}}{2m} \partial^{2}_{x} + \dfrac{1}{2} m \omega(t)^{2}(x-x_{0}(t))^{2} + g(t) |\psi(x,t)|^{2} \right) \psi(x,t).
\label{eq:relevant_GP}
\end{equation}
We wish to be able to extend the shortcut framework developed for the linear case. To do this we make the wavefunction ansatz for Eq. (\ref{eq:relevant_GP}) 
\begin{equation}
\psi(x,t) = e^{ - i \alpha_{2}(t) x^{2} + i \alpha_{1}(t) x - \beta(t) - i \mu \tau(t)} \phi \left( \dfrac{x - q_{c}(t)}{\rho(t)}\right) 
\label{eq:wavefn_ansatz}
\end{equation}
with $\tilde{x}= \dfrac{x-q_{c}(t)}{\rho(t)}$. Here $\phi(\tilde{x})$ is a solution of the stationary equation
\begin{equation}
\mu \phi(\tilde{x}) = - \dfrac{\hbar^{2}}{2m} \partial_{\tilde{x}}^{2}\phi(\tilde{x}) + \dfrac{m \omega_{0}^{2}}{2} \tilde{x}^{2} \phi(\tilde{x}) + g_{0} \left| \phi(\tilde{x}) \right|^{2} \phi(\tilde{x}).
\label{eq:GPE}
\end{equation}
Inserting Eq. (\ref{eq:wavefn_ansatz}) into Eq. (\ref{eq:relevant_GP}) we derive the following expression for $\alpha_{1}$ and $\alpha_{2}$
\begin{eqnarray}
\alpha_{1} &=& \dfrac{m}{\hbar \rho} (\dot{q}_{c}\rho-q_{c}\dot{\rho}),\\
\alpha_{2} &=& - \dfrac{m \dot{\rho}}{2 \rho \hbar}.
\end{eqnarray}
Additionally we obtain the following expression for $\tau$
\begin{eqnarray*}
\tau(t) = \dfrac{1}{2 \hbar \mu} \int_{0}^{t} \dfrac{1}{\rho^{2}(\tilde{t})} \bigg(2 \mu + m q_{c}^{2}(\tilde{t}) \dot{\rho}^{2}(\tilde{t}) -  2 m \rho(\tilde{t}) q_{c}(\tilde{t}) \dot{\rho}(\tilde{t}) \dot{q}_{c}(\tilde{t}) + m \rho^{2}(\tilde{t}) \dot{q}_{c}^{2}(\tilde{t})\\
 - m \rho(\tilde{t}) q_{c}^{2}(\tilde{t}) \ddot{\rho}(\tilde{t}) + m \rho^{2}(\tilde{t}) q_{c}(\tilde{t}) \ddot{q}_{c}(\tilde{t}) + m \rho^{2}(\tilde{t}) x_{0}(\tilde{t}) \ddot{q}_{c}(\tilde{t})\bigg) d\tilde{t}.
\end{eqnarray*}
We also recover the same auxiliary equations as in the single particle case
\begin{equation}
\rho^{3}(\ddot{\rho} + \rho \tilde{\omega}^{2})  - \tilde{\omega}_{0}^{2} = 0,
\end{equation}
\begin{equation}
\ddot{q}_{c} + \omega^{2}(q_{c}-x_{\text{min}}) = 0,
\end{equation}
and an additional equation for the atom-atom interaction
\begin{align*}
g(t) = \dfrac{g_{0}}{\rho(t)}.
\end{align*}
\end{widetext}

\end{document}